\documentclass[lnicst]{svmultln}

\usepackage{makeidx}  
\usepackage{graphicx}
\graphicspath{{figures/}}
\usepackage{url}
\urldef{\mailsc}\path|{vincent.taylor, ivan.martinovic}@cs.ox.ac.uk|

\begin{document}

\mainmatter              
\renewcommand{\arraystretch}{1.2}

\title{Quantifying Permission-Creep in the\\ Google Play Store}
\titlerunning{Quantifying Permission-Creep in the Google Play Store}

\author{Vincent F. Taylor \and Ivan Martinovic}
\authorrunning{Vincent F. Taylor and Ivan Martinovic}

\institute{Department of Computer Science,\\University of Oxford,\\Oxford, United Kingdom.\\ \mailsc\\}

\maketitle

\begin{abstract}
Although there are over~1,600,000 third-party Android apps in the Google Play Store, little has been conclusively shown about how their individual (and collective) permission usage has evolved over time. Recently, Android~6 overhauled the way permissions are granted by users, by switching to run-time permission requests instead of install-time permission requests. This is a welcome change, but recent research has shown that many users continue to accept run-time permissions blindly, leaving them at the mercy of third-party app developers and adversaries. Beyond intentionally invading privacy, highly privileged apps increase the attack surface of smartphones and are more attractive targets for adversaries. This work focuses exclusively on \textit{dangerous permissions}, i.e., those permissions identified by Android as guarding access to sensitive user data. By taking snapshots of the Google Play Store over a 20-month period, we characterise changes in the number and type of dangerous permissions used by Android apps when they are updated, to gain a greater understanding of the evolution of permission usage. We found that approximately~25,000 apps asked for additional permissions every three months. Worryingly, we made statistically significant observations that free apps and highly popular apps were more likely to ask for additional permissions when they were updated. By looking at patterns in dangerous permission usage, we find evidence that suggests developers may still be failing to correctly specify the permissions their apps need.

\keywords {Android; access control; permission evolution; Google Play Store}
\end{abstract}

\section{Introduction}

Android is the most popular mobile operating system with 84.7\% market share as of~2015~Q3, outpacing its nearest rival, iOS, at~13.1\%~\cite{marketshare}. This domination is fuelled by a myriad of app developers, devices, and consumers existing in a symbiotic relationship known as the Android ecosystem. Nielson reports that the average consumer uses over~26 different apps per month, spending more than one hour per day interacting with their smartphone~\cite{nielsonappusage}. This explosion in smartphone usage has been driven, in part, by the ease with which end-users can obtain third-party apps to extend the functionality of their devices. App marketplaces stand at the center of the ecosystem, acting as repositories for a plethora of apps, and providing convenient search and download facilities to satisfy the consumers' appetite. The Google Play Store is the largest (and only official) Android marketplace, boasting in excess of~1,600,000 apps~\cite{statistagoogleplay}.\\

\textbf{Run-Time Permissions.} The Android OS takes a permission-based approach to guard access to private user data and sensitive APIs. This approach requires users to explicitly accept (or reject) the permissions requested by an app. With the release of Android~6 (API level 23), users are no longer forced to accept (or reject) permissions in their entirety at install-time. Instead, users now accept (or reject) permissions individually at run-time. This offers added control, but many users continue to accept permission requests blindly due to conditioning or lack of understanding~\cite{Felt:2012:APU:2335356.2335360, Felt:2012:IG9:2381934.2381943}. This point is exemplified by Eling et al.~\cite{7427642} who show that 40.4\% of users continued to accept fine-grained, intrusive and unnecessary permission requests, in spite of the fact that these permission requests were \textit{presented to them at run-time}. Moreover, some app developers force the Android OS to revert to install-time permissions by making their apps compatible with older Android versions~\cite{backessoksp2016}, i.e., targeting API level 22 or lower. Thus run-time permissions, while helpful, are not the panacea they are hoped to be, as users' behaviour often diverges from their intentions when it comes to protecting their privacy~\cite{1392696, JOCA:JOCA70} and app developers can circumvent the new system. Even when there is no malicious intent from an app developer, apps with more permissions contribute a greater attack surface to the smartphone (in case of exploitable bugs/vulnerabilities) and any trend in app marketplaces concerning app (over)privilege must be thoroughly understood.\\

The permissions used by an app usually relate to the provision of the app's functionality, but some apps are intentionally permission-hungry, facilitating greater access to a user's personal data. Beyond app developers and advertisers (potentially) benefiting from this behaviour, smartphone security and privacy can be jeopardised if highly-privileged apps have vulnerabilities that can be exploited, for example, using permission-redelegation, app collusion, or confused deputy attacks~\cite{bugiel2012towards}. Throughout this paper, we focus exclusively on the so-called \textit{dangerous permissions} used in Android, i.e., those permissions that guard access to a user's confidential data~\cite{androiddevperms}. For this reason, we hereafter refer to \textit{dangerous permissions} as simply \textit{permissions}.

We are motivated to study app permission evolution, also known as permission-creep, across the entire Google Play Store, to quantify the effect that new versions of apps have on smartphone privacy/security as they are released into the app ecosystem. We also aim to understand the reasons for the addition/removal of permissions. By doing this, we aim to uncover micro- and macro-trends in the Android app ecosystem, which could potentially motivate future research directions. Previous research efforts have looked at permission evolution on the Android platform itself~\cite{wei2012permission}, or the evolution of permission usage in Android ad libraries~\cite{book2013longitudinal}, without looking at the changes in permission usage at the app level or across the entire Google Play Store. We consider this a critical gap in the literature and the underlying motivation for our work. Other work has looked at the Google Play Store to understand developer behaviour and pricing~\cite{Carbunar:2015:LSG:2808797.2808823} but neither at the breadth of our work (we analyse an order of magnitude more apps) nor to understand micro- or macro-trends.\\

\noindent{\textbf{Contributions.} Our paper makes the following contributions:}

\begin{itemize}
\item An evaluation of permission evolution across approximately~1,600,000 apps in the Google Play Store over a 20-month period.
\item An analysis of the number and types of permissions that are being added to (or removed from) apps, and how app attributes (such as cost or popularity) contribute to the likelihood of changes in permission usage.
\item An understanding of the reasons for permission additions/removals in apps.
\end{itemize}

\noindent{The rest of this paper is organised as follows: Section~2 surveys related work; Section~3 describes our dataset and the data collection methodology; Section~4 reports observations made from our dataset; Section~5 presents an analysis of permission usage and evolution across the dataset; Section~6 discusses our results, the limitations of our work, and future work; and finally Section~7 concludes the paper.}
\section{Related Work}

Viennot et al. performed the first large scale analysis of the Google Play Store using a tool they call PlayDrone to index and analyse over 1.1 million apps \cite{Viennot:2014:MSG:2591971.2592003}. They decompiled and obtained the source code for over 880,000 free apps. The authors characterised the content of the Google Play Store, measured library usage in apps, identified duplicate apps, and uncovered weaknesses in how authentication was implemented in many apps. Our work is similar in that we take snapshots of the entire Google Play store as well, but differs in that our analysis is longitudinal and is concerned with gaining a greater understanding of how app permission usage evolves over time and the potential impacts of this phenomenon on smartphone privacy and security.

More in line with our work, Book et al. do a longitudinal analysis of Android ad library permissions \cite{book2013longitudinal}. The authors investigate a sample of 114,000 apps to build a chronological map of permission usage in Android ad libraries. Their analysis reveals that indeed permission usage by ad libraries has increased in recent years and, worryingly, that these ad libraries have access to permissions that pose risks to user privacy and security. This work is a step in our direction, but since ad libraries request a subset of the permissions required by an app, it fails to capture the full picture of the risk to devices that come from apps on a whole. Our work expands by taking a holistic approach to looking at permission usage across the entire app.

Wei et al. go in a tangential direction and characterise permission evolution on the Android platform itself \cite{wei2012permission}. They look at changes in the Android permission model since its first commercial release for smartphones in 2008. They find that permission growth is aimed towards offering access to new hardware features and not towards offering more fine-grained control to the user. From a sample of 237 third-party apps, the authors discover that apps are over-privileged and use dangerous permissions. The authors also characterise the dangerous permissions required by pre-installed apps (which are difficult to remove) and argue that the evolution tends to favour the vendors and not the users. The authors focus mainly on permission evolution within the Android platform itself, while our work focuses on looking at permission evolution across all third-party apps in the official Android app market.

Along similar lines, Vidas et al. \cite{vidas2011curbing} propose a tool to help mitigate permission creep by assisting developers to enforce the principle of least privilege \cite{saltzer1975protection}. Their tool, implemented as an IDE plugin, analyses app source code and helps the developer to understand the minimum permissions required for their application to run. The authors focus on developers who inadvertently request too many permissions for reasons such as lack of understanding or expected future use. They create a permission-API database that understands the permissions required for particular function calls. From this database, their tool can statically analyse source code and manifest files to identify extraneous permissions. Using their tool, the authors examine 34,000 free third-party apps and find that more than 4\% have duplicate permissions in their manifest files. This underscores the idea that developers are not cautious enough when requesting permissions. Building on this, our work takes a look at whether there is a systematic permission creep across apps in the Google Play Store and whether conclusions can be drawn about the reason for this permission creep.

Finally, Carbunar and Potharaju~\cite{Carbunar:2015:LSG:2808797.2808823} analyse the Google Play Store to understand developer behaviour when it comes to publishing and pricing apps. They found that developers are more likely to increase the price of apps when they are updated. The authors also found that a few elite developers are responsible for very popular apps and that developers with many apps do not usually have any popular apps. This important work captures developer behaviour at a high level, i.e., publishing approaches, pricing, and app popularity. Complementary to this, we examine developer behaviour at a technical level, by analysing the security/privacy impact of app updates on the app ecosystem once they do happen.
\section{Dataset Description}

Collecting longitudinal data from a large source, such as the Google Play Store, is an arduous process. This is perhaps the reason that very few studies to date have focused on app permission usage, either at the magnitude (entire Google Play Store) or duration (over 1.5~years) that this one has. To gather our data, we designed a highly-scalable architecture that takes snapshots of the Google Play Store at fixed intervals.

\begin{figure}[!t]
\centering
\includegraphics[width=4.5in,clip=true,trim=0.24in 9.0in 3.4in 0.24in]{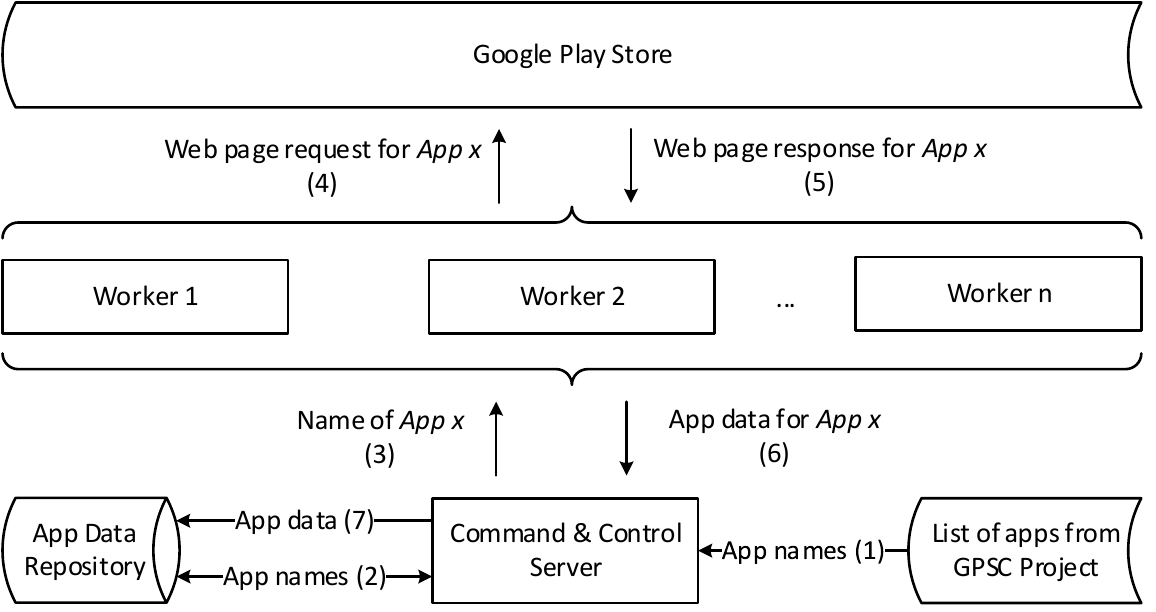}
\caption{Highly-scalable cloud-based crawler architecture. The Google Play Store Crawler (GPSC) Project~\cite{marcello} and Google Play Store~\cite{googleplaystore} itself were used as external sources of data. Numbers in brackets show the order of the steps in the process.}
\label{fig_crawler}
\end{figure}

Our long-term analysis of permission evolution requires data on all the apps in the Google Play Store. The Google Play Store Crawler (GPSC) project \cite{marcello} is concerned with exactly this, but unfortunately does not collect data on permission usage. Thus, we built our own crawler that retrieved full app data (including permission usage) from the Google Play Store, by leveraging the list of apps from the GPSC project. Our first snapshot of the Google Play Store (taken in this way) was taken in March 2015 with subsequent snapshots taken (at the same time of the month) every three months (quarterly) after the initial snapshot. Additionally, we leveraged a corpus of apps~\cite{archiveandroidapps} obtained using the PlayDrone tool~\cite{Viennot:2014:MSG:2591971.2592003}, to obtain an earlier snapshot of permission usage in the Google Play Store as at October 2014. The most recent snapshot used in our analysis is that of June 2016, for a total of seven snapshots (\texttt{October-2014}, \texttt{March-2015}, \texttt{June-2015}, \texttt{September-2015}, \texttt{December-2015}, \texttt{March-2016}, \texttt{June-2016}) covering a 20-month period. All snapshots are 3-months (quarterly) apart except the first two which are 5-months apart because we append the snapshot (\texttt{October-2014}) from prior work~\cite{archiveandroidapps}. We are careful to omit this first snapshot in some cases, such as when we quantify quarterly (3-month) changes across the Google Play Store.

In taking snapshots, our intention is to have the entire Store\footnote{We use the terms \textit{Google Play Store}, \textit{Store} and \textit{app store} interchangeably.} crawled as quickly as possible. To this end, we developed a cloud-based crawler, with geographically distributed worker nodes fetching app data and returning it to a command and control server. This is shown in Fig.~\ref{fig_crawler}. Our worker nodes make app store queries with random valid \textit{User-Agent} strings and are rate limited to \textit{3 requests/second}, to prevent blocking by the Google Play Store. Using a small-scale deployment with 3-4 workers, we can retrieve complete app data from the Google Play Store in less than 48 hours. Our most recent snapshot\footnote{Data from all our snapshots of the Google Play Store is available upon request.} of the Google Play Store, at the time of writing, contains 26.5GB of data on 2,003,739 apps available for download, and 728,445 apps that are no longer present in the Store.

With each new snapshot of the Google Play Store that we prepare to take, we carry-over the entire list of apps from the previous snapshot, and append any new apps that have been added to the store. Our system is informed of new additions to the Store from the GPSC project and our own crawlers. Apps that have been removed from the Google Play Store remain in our database, with an indicator that they are no longer available. Thus the number of apps in each of our snapshots monotonically increases as time progresses.
\section{Results}

\subsection{Overview of the Google Play Store}

\begin{figure}[]
\centering
\includegraphics[width=3.3in]{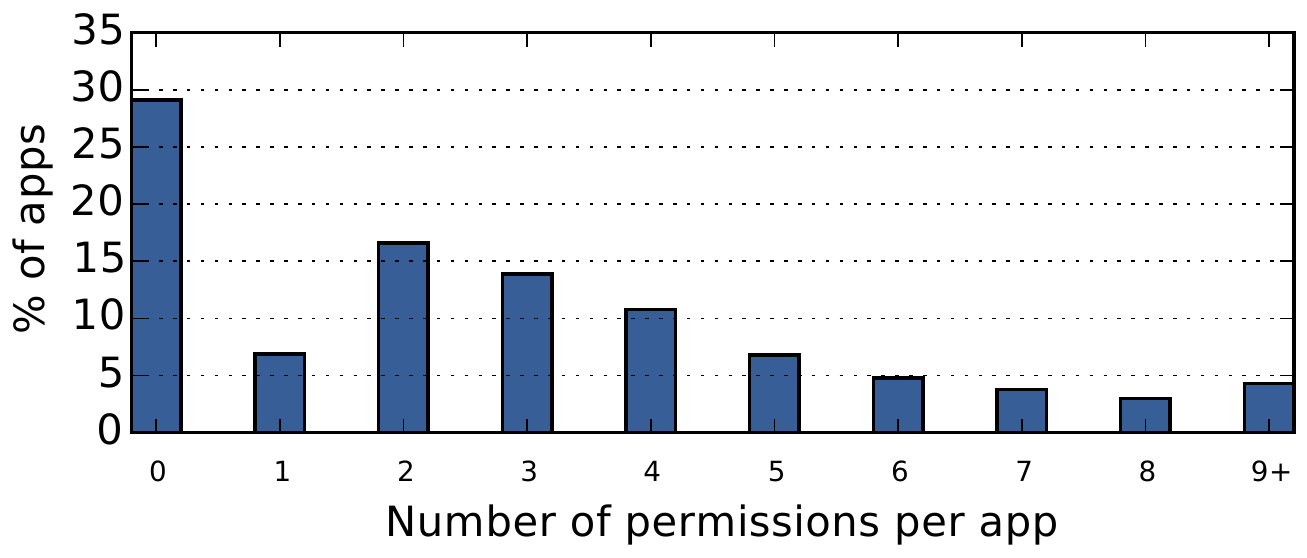}
\caption{Number of permissions used per app across the Google Play Store. Approximately 71\% of apps use one or more permissions.}
\label{fig_numperms-breakdown-entire-store}
\end{figure}

For generating the following overview of the Google Play Store, we leveraged our most current snapshot, i.e., \texttt{June-2016}. Table~\ref{table_perms-breakdown-entire-store} shows the popularity of permissions across all apps. The most popular permissions relate to reading from/writing to external storage on a device (59.3\%/58.7\% respectively), reading the current state of the device (33.7\%), and getting the user's fine/coarse location (25.3\%/24.3\%). Fig.~\ref{fig_numperms-breakdown-entire-store} shows the number of permissions used across all apps. The largest number of apps (29.1\%) used no permissions, while using two (16.6\%), three (13.9\%), and four (10.8\%) permissions was most common among apps that used permissions. Approximately 71\% of apps used one or more permissions.

\begin{table}[]
\centering
\caption{Prevalence of permission usage across the Google Play Store.}
\label{table_perms-breakdown-entire-store}
\begin{tabular}{p{2in}|p{0.5in}}
\hline
\texttt{READ\_EXTERNAL\_STORAGE} & 59.3\% \\ \hline
\texttt{WRITE\_EXTERNAL\_STORAGE} & 58.7\% \\ \hline
\texttt{READ\_PHONE\_STATE} & 33.7\% \\ \hline
\texttt{ACCESS\_FINE\_LOCATION} & 25.3\% \\ \hline
\texttt{ACCESS\_COARSE\_LOCATION} & 24.3\% \\ \hline
\texttt{GET\_ACCOUNTS} & 23.6\% \\ \hline
\texttt{CAMERA} & 15.8\% \\ \hline
\texttt{RECORD\_AUDIO} & 8.74\% \\ \hline
\texttt{CALL\_PHONE} & 8.43\% \\ \hline
\texttt{READ\_CONTACTS} & 6.14\% \\ \hline
\texttt{SEND\_SMS} & 3.63\% \\ \hline
\texttt{WRITE\_CONTACTS} & 3.05\% \\ \hline
\texttt{RECEIVE\_SMS} & 2.45\% \\ \hline
\texttt{READ\_CALL\_LOG} & 2.45\% \\ \hline
\texttt{READ\_CALENDAR} & 2.18\% \\ \hline
\texttt{WRITE\_CALL\_LOG} & 1.45\% \\ \hline
\texttt{WRITE\_CALENDAR} & 1.35\% \\ \hline
\texttt{READ\_SMS} & 1.24\% \\ \hline
\texttt{PROCESS\_OUTGOING\_CALLS} & 0.923\% \\ \hline
\texttt{RECEIVE\_MMS} & 0.128\% \\ \hline
\texttt{USE\_SIP} & 0.118\% \\ \hline
\texttt{BODY\_SENSORS} & 0.017\% \\ \hline
\texttt{RECEIVE\_WAP\_PUSH} & 0.008\% \\ \hline
\texttt{ADD\_VOICEMAIL} & 0.0007\% \\ \hline
\end{tabular}
\end{table}


\begin{figure}[]
\centering
\includegraphics[width=3.3in]{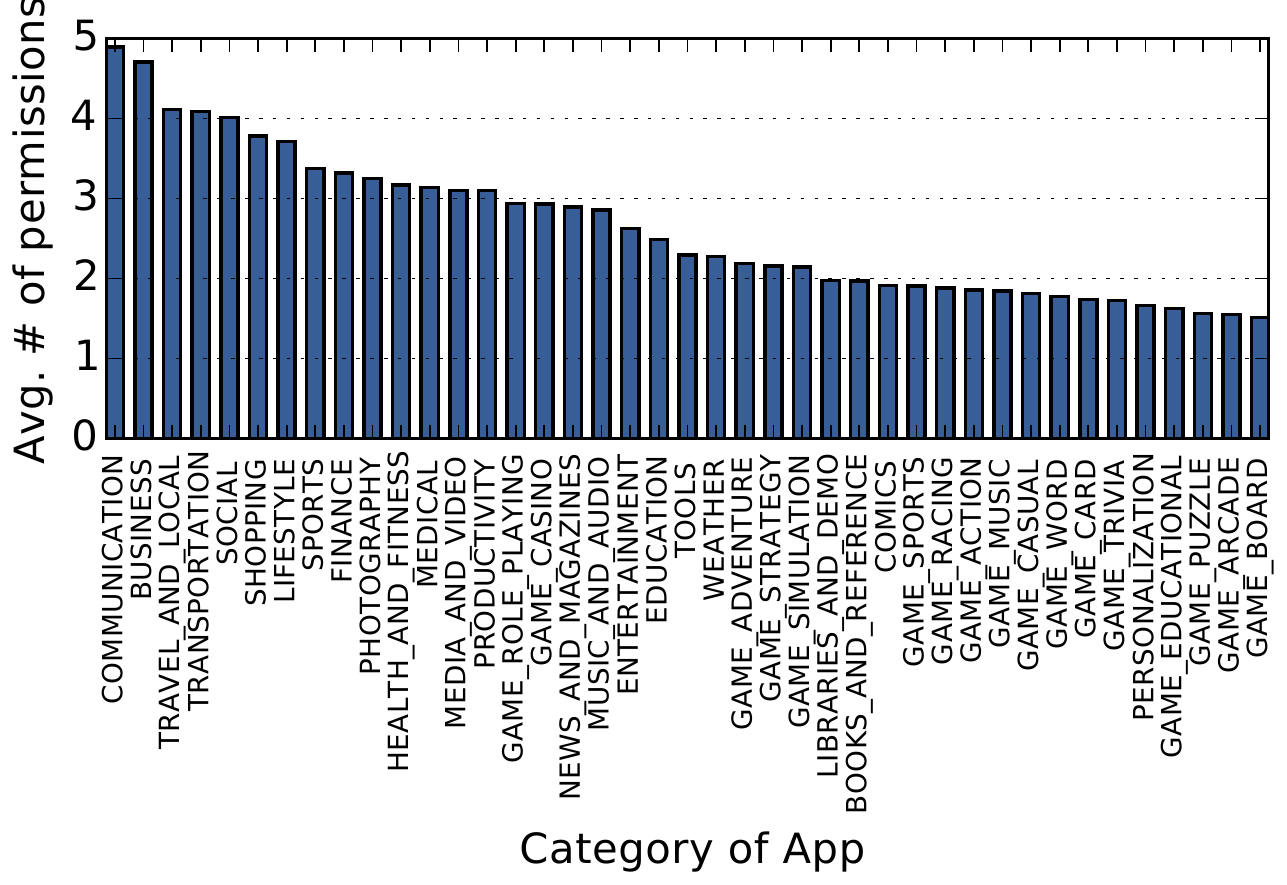}
\caption{Average number of permissions used per category of app.}
\label{fig_numperms-per-category}
\end{figure}

\begin{figure}[]
\centering
\includegraphics[width=3.3in]{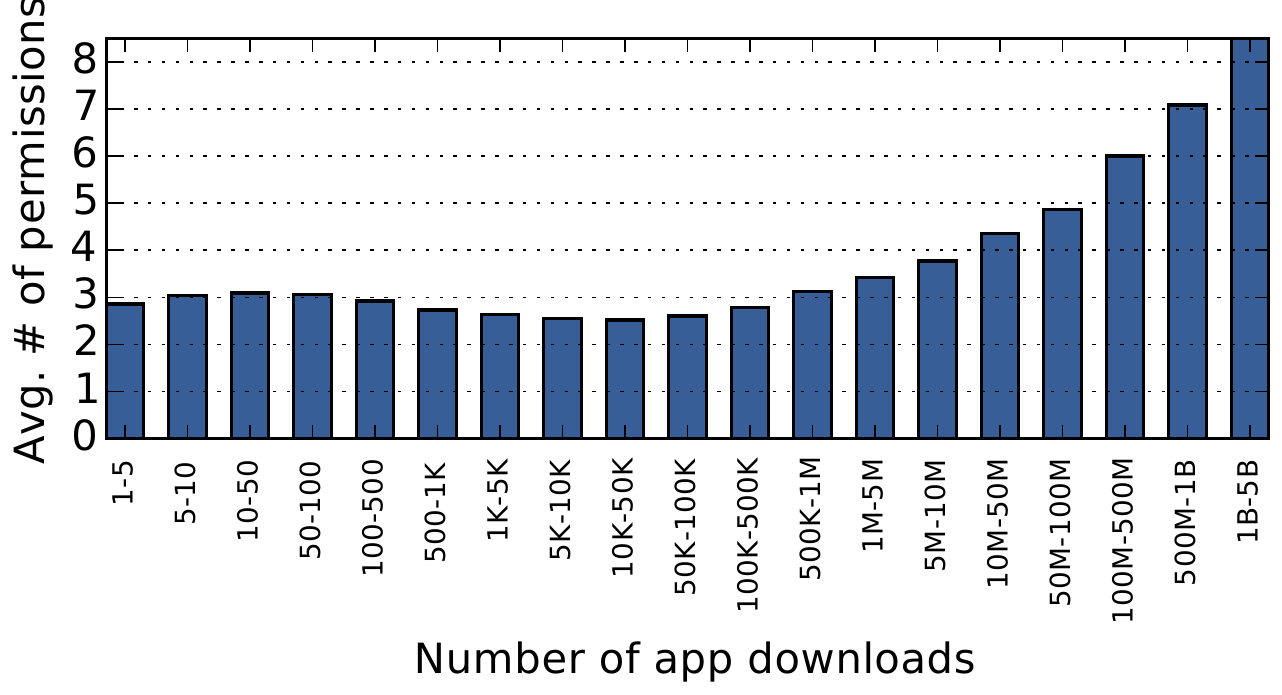}
\caption{Average number of permissions used per number of downloads of app. K = thousand, M = million, B = billion.}
\label{fig_perms-per-number-of-downloads}
\end{figure}

Fig.~\ref{fig_numperms-per-category} shows the average number of permissions used by apps based on the category that they were listed under. \mbox{\texttt{COMMUNICATION}} and \texttt{BUSINESS} were the most permission-hungry categories, using an average of 4.9 and 4.7 permissions respectively. The other end of the spectrum was predominantly games, with the overall least permission-hungry category of app being \texttt{GAME\_BOARD} using an average of 1.5 permissions per app. Fig.~\ref{fig_perms-per-number-of-downloads} shows how many permissions were used based on the number of downloads that an app had. Up to \texttt{1M} downloads, the average app used approximately 2.5-3 permissions. Above \texttt{10K} downloads, apps had permission usage that monotonically increased from 2.5 to 8.5 permissions for the most popular apps (\texttt{1B} downloads or more).

\subsection{Permission Changes Over Time}
\label{subsection_app-updates}

\begin{figure}[]
\centering
\includegraphics[width=3.3in]{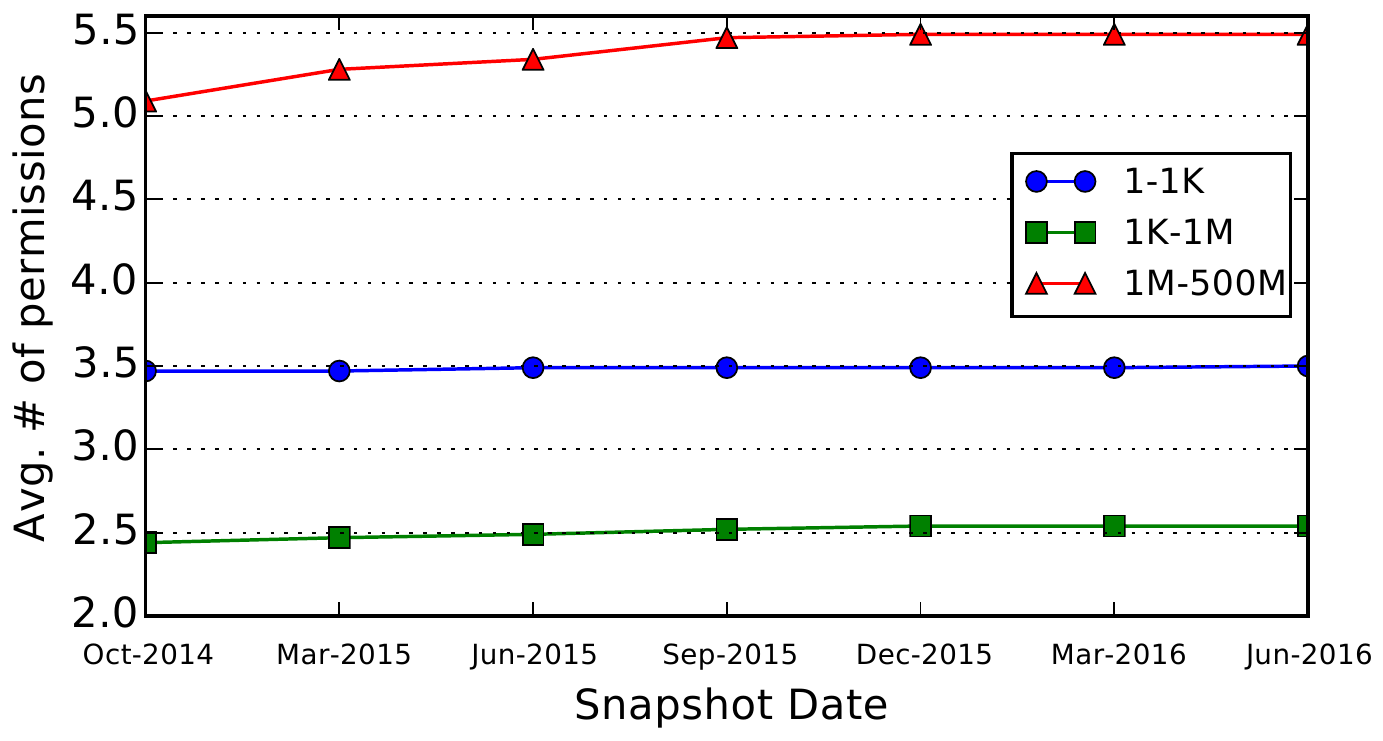}
\caption{Permission usage over time by number of downloads. Less downloaded apps had a greater increase in the number of permissions used.}
\label{fig_longitudinal-per-number-of-downloads}
\end{figure}

Fig.~\ref{fig_longitudinal-per-number-of-downloads} shows how permission usage changed across the Google Play Store over the 20-month period of \texttt{October-2014} to \texttt{June-2016}. For this analysis, we used only those apps that were in all snapshots over the entire period. We analysed permission usage based on the number of downloads of an app. We divided apps into three categories based on their total number of downloads: \texttt{\mbox{1-1K}} (low downloads), \texttt{\mbox{1K-1M}} (medium downloads), and \texttt{\mbox{1M-5B}} (high downloads). Across all snapshots, every category of app had an increase in the number of permissions used. Overall, we found that apps in the \textit{high downloads} category had the highest increase in permission usage over the 20-month period, going from 5.1 to 5.5 permissions on average. Apps in the \textit{medium downloads} category went from using 2.44 to 2.54 permissions on average. Apps in the \textit{low downloads} category had the lowest increase in permission usage, going from 3.47 to 3.50 over the studied period. In terms of absolute change in the number of permissions used, more popular apps had the greater increase. While the absolute change in mean permission usage overall seems small, it is important to remember that these numbers reflect the aggregate permission change across all apps in the Google Play Store, including those apps that were not updated at all. At the granularity of individual apps, addition of several permissions between snapshots is not uncommon, as we show later in this section.

We analysed permission increase/decrease at the app level to understand how many permissions individual apps were adding or removing when they were updated\footnote{We used a change in an app's version or file size (as reported in the Google Play Store) between one snapshot and the subsequent snapshot as a proxy to determine whether an app was updated.}. We observed that a majority (86\%) of apps did not get updated between snapshots of the Google Play Store. In the following analysis, we include only those apps that were updated, to understand what changes in permission usage, if any, were made when apps were updated. Table~\ref{table-permission-change} shows the breakdown of permission changes across our quarterly snapshots for those apps that were updated. In the table, each date shows the permission changes between that snapshot and the previous snapshot. Note that we omit the \texttt{October-2014} snapshot in this analysis because it was 5-months from its subsequent snapshot instead of 3-months. The table details the amount of permissions that were added/removed across the apps. From the table, the majority of apps (on average 78.9\%) did not have any change in permission usage between snapshots. For apps that did have changes, the most likely change (7.1\% of apps on average) was to add one new permission (+1). Averaging permission changes across our 20-months of data, 13.2\% (approximately 25,000) of apps added one or more new permissions every 3-months, while only 7.9\% of apps removed one or more permissions. That is, approximately twice (1.7) as many apps required an increased number of permissions than those that required a decrease, over the studied period.

\begin{table}[]
\centering
\caption{Table showing how permission usage changed between quarters for those apps that were updated. Approximately twice as many apps added permissions than removed permissions.}
\label{table-permission-change}
\begin{tabular}{l|l|l|l|l|l}

 & \textbf{Jun-15} & \textbf{Sep-15} & \textbf{Dec-15} & \textbf{Mar-16} & \textbf{Jun-16} \\ \hline
\textbf{Total Increase} & \textbf{11.82\%} & \textbf{16.23\%} & \textbf{13.82\%} & \textbf{12.50\%} & \textbf{11.66\%} \\ \hline
+6 or More & 0.24\% & 0.25\% & 0.23\% & 0.21\% & 0.23\% \\ \hline
+5         & 0.15\% & 0.22\% & 0.25\% & 0.19\% & 0.13\% \\ \hline
+4         & 0.55\% & 0.80\% & 0.67\% & 0.55\% & 0.40\% \\ \hline
+3         & 1.25\% & 1.50\% & 1.45\% & 1.20\% & 1.12\% \\ \hline
+2         & 3.21\% & 4.65\% & 4.03\% & 3.50\% & 3.50\% \\ \hline
+1         & 6.42\% & 8.81\% & 7.19\% & 6.85\% & 6.28\% \\ \hline \hline
\textbf{No Change} & \textbf{82.19\%} & \textbf{78.20\%} & \textbf{77.21\%} & \textbf{78.27\%} & \textbf{78.51\%} \\ \hline \hline
-1         & 3.03\% & 2.95\% & 5.72\% & 5.16\% & 4.84\% \\ \hline
-2         & 1.82\% & 1.45\% & 1.62\% & 2.31\% & 3.19\% \\ \hline
-3         & 0.53\% & 0.56\% & 0.99\% & 1.03\% & 1.04\% \\ \hline
-4         & 0.21\% & 0.22\% & 0.34\% & 0.39\% & 0.40\% \\ \hline
-5         & 0.12\% & 0.11\% & 0.12\% & 0.13\% & 0.14\% \\ \hline
-6 or Less & 0.28\% & 0.28\% & 0.18\% & 0.21\% & 0.22\% \\ \hline
\textbf{Total Decrease} & \textbf{5.99\%} & \textbf{5.57\%} & \textbf{8.97\%} & \textbf{9.23\%} & \textbf{9.83\%}  \\
\end{tabular}
\end{table}

\begin{figure}[]
\centering
\includegraphics[width=3.3in]{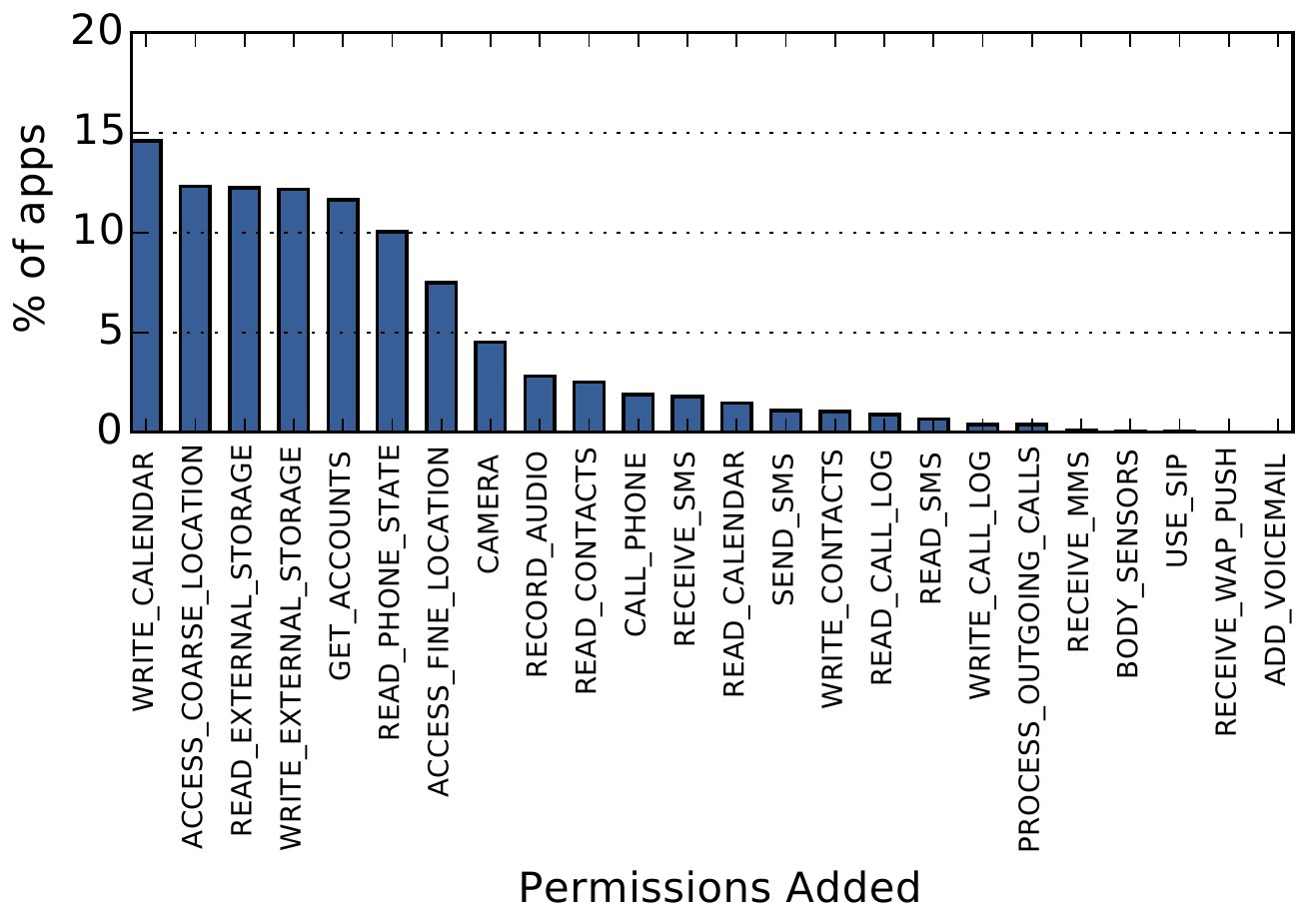}
\caption{Breakdown of the newly added permissions across the Google Play Store.}
\label{fig_which-permissions-were-added-most}
\end{figure}

We analysed the permissions that were added to apps when they were updated, to understand the potential erosion in privacy caused. This result is presented in Fig.~\ref{fig_which-permissions-were-added-most}. From the figure, we can see that the Top~5 permissions that were added were \texttt{WRITE\_CALENDAR}, \texttt{ACCESS\_COARSE\_LOCATION}, \texttt{READ\_EXTERNAL\_STORAGE, WRITE\_EXTERNAL\_STORAGE}, and \texttt{GET\_ACCOUNTS}. These permissions allow an app access to write to a user's calendar events, access their general location, read from and write to their external storage, and access the list of accounts (in the Accounts Service) on the user's device~\cite{manifestperms}. Android automatically grants new permissions if the user has already accepted a permission from the same permission group~\cite{androiddevperms}. Thus, the Top~5 added permissions also allow an app to read a user's calendar and get their precise location. In general, the most commonly added permissions allowed apps to read additional sensitive data from a device.

\begin{figure}[]
\centering
\includegraphics[width=3.3in]{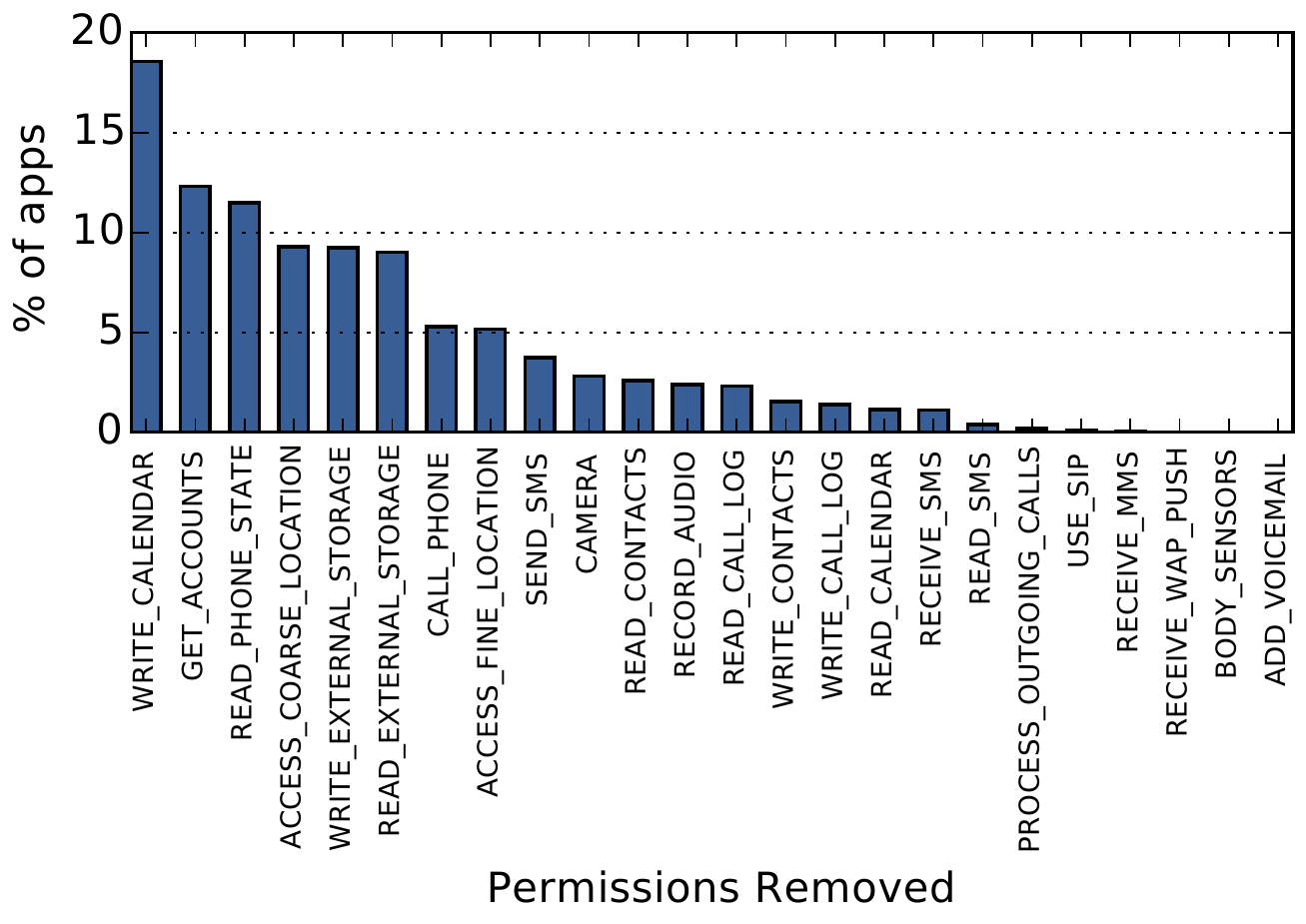}
\caption{Breakdown of the newly removed permissions across the Google Play Store.}
\label{fig_which-permissions-were-removed-most}
\end{figure}

Fig.~\ref{fig_which-permissions-were-removed-most} shows which permissions were removed the most. The most commonly removed permission was \texttt{WRITE\_CALENDAR}, covering approximately 18\% of the incidents of permission removal from apps. It is interesting to note that \texttt{WRITE\_CALENDAR} was both the most frequently added and most frequently removed permission. This may be due to developer confusion about the necessity of the particular permission. Such developer confusion regarding permission usage was also observed by Felt et al.~\cite{Felt:2011:APD:2046707.2046779}.

\section{Analysis}

\subsection{Patterns in permission addition and removal}

Our findings so far have examined aggregate permission evolution at the macro-level. We now use a technique similar to Wei et al.~\cite{wei2012permission}, to understand permission evolution at the micro-level. Specifically, we leverage patterns to better understand permission addition and removal by developers at the app level. Patterns are written using the digits \texttt{0} and \texttt{1} where a \texttt{0} represents the state that an app does not use a permission and \texttt{1} represents the state where an app uses a permission. Transitions between states are represented using a $\rightarrow$. Thus, the simplest event where an app goes from not using a permission to using a permission can be illustrated as \texttt{0}$\rightarrow$\texttt{1}. Using this idea of permission state transitions, we created what we call Permission Evolution Matrices (PEMs) by looking at permission usage in apps across snapshots of the Google Play Store. Using PEMs we can easily identify patterns in permission evolution. An example of the PEM for an app is shown in Table~\ref{fig_app-permission-matrix}.

\begin{table}[]
\centering
\caption{Example of an (abridged) permission evolution matrix (PEM). The pattern \texttt{0}$\rightarrow$\texttt{1}$\rightarrow$\texttt{0}$\rightarrow$\texttt{1} for the \texttt{READ\_PHONE\_STATE} permission can be seen.}
\label{fig_app-permission-matrix}
\begin{tabular}{rcccccc}
Permission & Mar-15 & Jun-15 & Sep-15 & Dec-15 & Mar-16 & Jun-16 \\ \hline
\texttt{READ\_PHONE\_STATE} & 0 & \textbf{0} & \textbf{1} & \textbf{0} & \textbf{1} & 1 \\
\texttt{SEND\_SMS} & 1 & 1 & 1 & 1 & 1 & 1 \\
\texttt{READ\_SMS} & 1 & 1 & 1 & 1 & 1 & 1 \\
\texttt{CALL\_PHONE} & 0 & 0 & 0 & 0 & 0 & 0 \\
\texttt{GET\_ACCOUNTS} & 0 & 0 & 0 & 0 & 0 & 0 \\
... & ... & ... & ... & ... & ... & ... \\
\texttt{READ\_CONTACTS} & 1 & 1 & 1 & 1 & 1 & 1
\end{tabular}
\end{table}

We evaluated the PEMs for all apps looking for interesting patterns such as \texttt{0}$\rightarrow$\texttt{1}$\rightarrow$\texttt{0}. We consider the pattern \texttt{0}$\rightarrow$\texttt{1}$\rightarrow$\texttt{0} to be interesting because it means that an app without a particular permission first added the permission, then removed it shortly after. This is strange behaviour and may suggest app developer confusion regarding the necessity of the permission in question. This could also be caused by the requirements of the libraries used by an app, however we consider this unlikely since, from observation, libraries tend to have constant (or increasing) permission usage, but an oscillation is unusual.

\begin{table}[]
\centering
\caption{Number of incidences of apps with the pattern \texttt{0}$\rightarrow$\texttt{1}$\rightarrow$\texttt{0}.}
\label{fig_app-permission-pattern-010}
\begin{tabular}{p{2in}|l}
Permission & Number of incidences \\ \hline
\texttt{WRITE\_CALENDAR} & 7574 \\ \hline
\texttt{ACCESS\_COARSE\_LOCATION} & 3151 \\ \hline
\texttt{GET\_ACCOUNTS} & 1433 \\ \hline
\texttt{WRITE\_EXTERNAL\_STORAGE} & 977 \\ \hline
\texttt{READ\_EXTERNAL\_STORAGE} & 954 \\ \hline
\texttt{READ\_PHONE\_STATE} & 916 \\ \hline
\texttt{ACCESS\_FINE\_LOCATION} & 558 \\ \hline
\texttt{RECORD\_AUDIO} & 305 \\ \hline
\texttt{READ\_CONTACTS} & 210 \\ \hline
\texttt{CAMERA} & 178 \\ \hline
\texttt{READ\_CALL\_LOG} & 165 \\ \hline
\texttt{CALL\_PHONE} & 152 \\ \hline
\texttt{WRITE\_CONTACTS} & 89 \\ \hline
\texttt{READ\_CALENDAR} & 73 \\ \hline
\texttt{RECEIVE\_SMS} & 70 \\ \hline
\texttt{SEND\_SMS} & 57 \\ \hline
\texttt{WRITE\_CALL\_LOG} & 54 \\ \hline
\texttt{READ\_SMS} & 48 \\ \hline
\texttt{PROCESS\_OUTGOING\_CALLS} & 28 \\ \hline
\texttt{RECEIVE\_MMS} & 5 \\ \hline
\texttt{ADD\_VOICEMAIL} & 2 \\ \hline
\texttt{USE\_SIP} & 2 \\ \hline
\texttt{RECEIVE\_WAP\_PUSH} & 2 \\ \hline
\texttt{BODY\_SENSORS} & 1 \\ \hline
\end{tabular}
\end{table}

Table~\ref{fig_app-permission-pattern-010} shows incidences of apps with the pattern \texttt{0}$\rightarrow$\texttt{1}$\rightarrow$\texttt{0}. The most common oscillating permissions that \textit{read} data from a smartphone were \texttt{GET\_ACCOUNTS}, \texttt{ACCESS\_COARSE\_LOCATION}, and \texttt{READ\_EXTERNAL\_STORAGE}. These permissions allow an app to get a user's general location, get the names/email addresses of accounts on the device, and read files located on external storage such as an SD card. The existence of the pattern \texttt{0}$\rightarrow$\texttt{1}$\rightarrow$\texttt{0} for an app suggests that the app may have been over-privileged for a period of time before the developers corrected the error, since we deem it unlikely that developers would be adding and removing features (needing permissions) so rapidly. Also, we consider the number of occurrences of patterns reported here to be a lower bound, because the granularity of our snapshots was 3-months. Indeed, more granular snapshots may have revealed more ephemeral patterns.

We further analysed the apps that had the interesting permission transition pattern \texttt{0}$\rightarrow$\texttt{1}$\rightarrow$\texttt{0}. Overwhelmingly, 96.5\% of these apps were free apps and 81\% of them had less than 50,000 downloads. Thus, it seems that fledgling apps (and their developers) disproportionately suffer from the permission usage oscillations that we discovered, again pointing to lack of understanding or expertise. Note however, that it was not only fledgling apps that had trouble with permission usage oscillations. Indeed, nine apps with \texttt{100M-500M} downloads had oscillating permissions. For example, \texttt{com.avast.android.mobilesecurity} briefly used \texttt{ACCESS\_COARSE\_LOCATION}, while \texttt{com.ijinshan.kbatterydoctor\_en} and \texttt{vStudio.Android.Camera360} briefly used the \texttt{GET\_ACCOUNTS} permission.

Permission usage information is directly extracted from an app's manifest file by the Google Play Store. Thus, an app may not actually use a listed permission internally; it may have only been added to the package manifest. In any case, this is cause for concern because it is a sign of developer error. Moreover, even if this permission was not used by the app, it would still increase the attack surface of the app and smartphone if, for example, the app had vulnerabilities or used dynamic code loading (or other low-level functionality) that could be exploited by an adversary. Also, apps asking for unnecessary permissions condition users to accept many permission requests, which is already a problem. As mentioned above, even very popular apps suffer from this problem of fluctuating permissions, potentially affecting a non-trivial portion of the ecosystem.

\subsection{Impact of the cost of an app}

We examined the impact, if any, that the cost (free or paid) of an app had on its likelihood to add new permissions over time. Table~\ref{table-permission-increase-incidents-free-vs-paid} shows a breakdown of incidents where permissions were added to apps based on their cost.\\

\begin{table}[]
\centering
\caption{Aggregated list of incidents where permissions were added, broken down by whether an app was free or paid.}
\label{table-permission-increase-incidents-free-vs-paid}
\begin{tabular}{|l|l|l|l|l|}
\hline
\textbf{Increase} & \textbf{Free} & \textbf{Free (\%)} & \textbf{Paid} & \textbf{Paid (\%)} \\ \hline
1 & 49858 & 95.7\% & 2233 & 4.3\% \\ \hline
2 & 20011 & 94.6\% & 1142 & 5.4\% \\ \hline
3 & 6914 & 96.2\% & 276 & 3.8\% \\ \hline
4 & 3387 & 96.8\% & 111 & 3.2\% \\ \hline
5 & 1071 & 97.3\% & 30 & 2.7\% \\ \hline
6 & 473 & 97.1\% & 14 & 2.9\% \\ \hline
7 & 229 & 99.6\% & 1 & 0.4\% \\ \hline
8 & 122 & 98.4\% & 2 & 1.6\% \\ \hline
9 & 75 & 72.8\% & 28 & 27.2\% \\ \hline
10+ & 164 & 62.6\% & 98 & 37.4\% \\ \hline
\textbf{Total} & \textbf{82304} & \textbf{95.4\%} & \textbf{3935} & \textbf{4.6\%} \\ \hline
\end{tabular}
\end{table}

\noindent{\textbf{Hypothesis 1: Free apps are more likely to add new permissions than paid apps.}\\\\We wanted to determine whether there was a statistically significant difference between free vs. paid apps when it came to adding new permissions. We tested this using a 2-proportion z-test with a sample size of 20,000, at a significance level of 0.01. The data from our random sample of the dataset is shown in Table~\ref{table-cost-ztest-random-sample}. Our result showed $p < 0.01$, suggesting that free apps were more likely than their paid counterparts to add new permissions over time. This result, perhaps seemingly intuitive and unsurprising to some, has not been shown before in the literature. Free apps tend to be ad-supported and thus could be adding more permissions to satisfy the needs of advertising libraries. If so, this is cause for concern, and underscores the need for privilege separation between apps and their ad libraries~\cite{180236}.}

\begin{table}[]
\centering
\caption{Results of the random sample ($n = 20,000$) of apps based on their cost.}
\label{table-cost-ztest-random-sample}
\begin{tabular}{|l|c|c|}
\hline
 & \textbf{Total Apps} & \textbf{Apps adding permissions} \\ \hline
\textbf{Free} & 17913 & 861 \\ \hline
\textbf{Paid} & 2087 & 44 \\ \hline
\end{tabular}
\end{table}

\subsection{Impact of the popularity of an app}

Next, we examined the impact that the popularity (number of downloads) of an app had on permission usage over time. Table~\ref{table-permission-increase-incidents-num-downloads} breaks down incidents\footnote{The discrepancy of 58 between the total number of incidents in Table~\ref{table-permission-increase-incidents-free-vs-paid} and Table~\ref{table-permission-increase-incidents-num-downloads} is caused by errors in the data obtained from the Google Play Store. These errors prevent us from accurately parsing the number of downloads for 58 apps, and thus these apps have been omitted.} of permissions being added, based on app popularity (whether an app had Low or High downloads).\\

\noindent{\textbf{Hypothesis 2: Popular apps are more likely to add new permissions than less popular apps.}\\\\ As in Section~\ref{subsection_app-updates}, we considered 1-million downloads as the threshold above which an app was placed in the High category. We used a 2-proportion z-test with a sample size of 20,000, and significance level of 0.01 to determine whether there was a statistically significant difference between apps with Low and High downloads when it came to adding new permissions. The data from our random sample of the dataset is shown in Table~\ref{table-popularity-ztest-random-sample}. Our result showed $p < 0.01$, suggesting that apps in the High downloads category were more likely to add new permissions over time. This result has also never been shown in the literature. It is somewhat concerning that very popular apps are adding new permissions to a greater extent than less popular apps. This is because it means that a large cross-section of users (those using very popular apps) will use apps with a larger attack surface. We expect that the developers of very popular apps take user privacy and security seriously and this mitigates the risk somewhat. In other worrying cases, some very popular apps (e.g. flashlight, alarm clock) are developed by small teams or individuals who may not have the knowledge, desire or capacity to write their apps according to best practices, and increase the risk to users when their apps leverage additional permissions. Additionally, if very popular apps are the ones most likely to use new permissions, it means that large swathes of users face the risk of being conditioned to automatically accept permission requests.}

\begin{table}[]
\centering
\caption{Aggregated list of incidents where permissions were added, broken down by number of downloads. Low refers to apps having less than 1-million downloads while High refers to apps with more than 1-million downloads.}
\label{table-permission-increase-incidents-num-downloads}
\begin{tabular}{|l|l|l|l|l|}
\hline
\textbf{Increase} & \textbf{Low} & \textbf{Low (\%)} & \textbf{High} & \textbf{High (\%)} \\ \hline
1 & 49719 & 95.5\% & 2321 & 4.5\% \\ \hline
2 & 20200 & 95.6\% & 938 & 4.4\% \\ \hline
3 & 6869 & 95.5\% & 321 & 4.5\% \\ \hline
4 & 3373 & 96.4\% & 125 & 3.6\% \\ \hline
5 & 1047 & 95.5\% & 49 & 4.5\% \\ \hline
6 & 464 & 95.5\% & 22 & 4.5\% \\ \hline
7 & 236 & 96.7\% & 8 & 3.3\% \\ \hline
8 & 115 & 92.7\% & 9 & 7.3\% \\ \hline
9 & 101 & 98.1\% & 2 & 1.9\% \\ \hline
10+ & 257 & 98.1\% & 5 & 1.9\% \\ \hline
\textbf{Total} & \textbf{82381} & \textbf{95.6\%} & \textbf{3800} & \textbf{4.4\%} \\ \hline
\end{tabular}
\end{table}

\begin{table}[]
\centering
\caption{Results of the random sample ($n = 20,000$) of apps based on their popularity.}
\label{table-popularity-ztest-random-sample}
\begin{tabular}{|l|c|c|}
\hline
 & \textbf{Total Apps} & \textbf{Apps adding permissions} \\ \hline
\textbf{Low Downloads} & 19534 & 905 \\ \hline
\textbf{High Downloads} & 202 & 37 \\ \hline
\end{tabular}
\end{table}
\section{Discussion}

Our focus throughout this paper was to understand how permission usage in the Android ecosystem evolves over time. While run-time permission requests put more power into the hands of the end-user, Eling et al.~\cite{7427642} show that a substantial portion of users still blindly accept permission requests when they are confronted with them. Thus, it is important to understand the current state of the app ecosystem (in terms of permission usage by apps) and where it may likely progress to in the future. We were interested in understanding whether the cost (free vs. paid) or popularity (number of downloads) of an app had any bearing on the usage/evolution of permissions. We were also interested in detecting interesting phenomena, such as permission usage patterns, that could signify developer error in specifying permissions. We only focused on \textit{dangerous permissions}, those permissions defined by the Android operating system as guarding sensitive user data. We ignored changes in \textit{normal permissions} because these permissions, if abused, only cause minor annoyance to a user, as opposed to putting their personal data at risk.

The addition of new permissions is not inherently bad, as many apps have legitimate reasons to request additional access to user data. However, many apps and ad libraries are also known to abuse their granted permissions for the purposes of profiling users and/or directly stealing their data. Regardless of the intent of the app developer when adding new permissions, more highly-privileged apps contribute to a greater attack surface on the smartphone as well as attract the attention of adversaries. 

We first looked at the average number of permissions used by apps across the Google Play Store based on the popularity of an app. We found that very popular apps (those with in excess of one million downloads) used more permissions than those with less downloads. This could be due to the fact that they provide more functionality, and hence why they are popular in the first place. Thus popular apps may be more attractive to adversaries since they offer a greater attack surface and are more widely installed. Looking at apps that were present across all our snapshots, we observed that more popular apps had a greater overall increase in average permission usage.

We attempted to understand the frequency of app updates and the extent to which permissions were added (or removed) when apps were updated. Unsurprisingly, we observed that the majority of apps (86\%) did not get updated between Store snapshots. Of the apps that did get updated, approximately four out of five of them did not have any change in their permission requirements. However, 13.2\% of the apps that were updated now needed one or more new permissions. This corresponds to approximately 25,000 apps needing new permissions every three months. We discovered that apps were more likely to add new permissions than remove permissions, overall pointing to greater access to personal data by apps. This can obviously contribute to an erosion in privacy, but the security impacts also need to be considered since highly-privileged apps are more attractive targets to adversaries.

We looked at patterns in permission usage as apps evolved across our dataset. Specifically, we looked for oscillations in permission usage, as this can be considered strange behaviour. We found that free apps and those with fewer downloads disproportionally suffered from this phenomenon, potentially pointing to lack of developer expertise or confusion about what permissions were actually needed. We also found that some popular apps suffered from this problem, showing that no set of apps/app developers is immune.

We conducted hypothesis tests to determine whether the cost of an app had any impact on its likelihood to add new permissions. We found statistically significant evidence that free apps were more likely to add new permissions than paid apps. This could be as a result of the advertising libraries that are usually included with free apps, which leverage personal data to provide more targeted advertisements. Whether it is the app or the ad library that leverages these new permissions, the fact remains that there is now additional opportunity for user data to be abused. We tested whether the popularity of an app had an impact on how likely the app was to add new permissions. Our data showed that more popular apps were more likely to add new permissions over. This is worrying, because it means that a large segment of users in the Android ecosystem are potentially exposed to additional privacy/security risks simply as a result of using and updating popular apps.

\subsection{Limitations}

In this work, we took snapshots of the Google Play Store at 3-month intervals. While this is useful for getting an overall picture of the Store, it may not be granular enough to capture short-term phenomena when they happen. In looking at permission usage, we are relying on the information reported in the Google Play Store regarding the permissions listed in an app's manifest. There is no way to easily disaggregate and understand whether the app or its included libraries were responsible for using these permissions. However, regardless of whether it is the app or libraries using the permission, high-privilege in apps increases the attack surface for that app and the smartphone it is installed on, as well as attracts the attention of adversaries. Moreover, since privileges are not separated between apps and their libraries, either is free to leverage the access it has been granted at the expense of the user.

\subsection{Future Work}

For future work, we plan to continue the collection of our snapshots of data as well as increase the frequency of snapshot collection. We also plan to look at additional metrics beyond permission usage that would allow a more comprehensive understanding of how the app ecosystem is evolving.
\section{Conclusion}

In this paper, we did a longitudinal analysis of permission evolution in apps by taking quarterly snapshots of the Google Play Store over a 20-month period. Our analysis showed that the average number of permissions used by apps increased over the period, with more popular apps having a greater increase. Of the apps that do have changes in their permission requirements, the most likely change was the addition of a new permission. We observed that almost twice as many apps added new permissions as those that removed permissions. We did hypothesis testing and observed that free apps and popular (apps with more than one million downloads) apps were more likely to add new permissions over time. We also found oscillation in permission usage, which suggests that app developers may still be confused about what permissions their apps require. This has the potential to make apps over-privileged and unnecessarily increase the attack surface. By performing this longitudinal study, we have confirmed and highlighted several important trends in app permission usage across the Google Play Store; never before done in the literature. By drawing these trends to the attention of the research community, we hope to generate additional interest in the area, so that appropriate strategies can be developed to keep sensitive user data safe, as smartphones continue their growth to ubiquity.

\section{Acknowledgments}
Vincent F. Taylor is supported by a Rhodes Scholarship and the UK EPSRC. Thanks to Marcello Lins of the Google Play Store Crawler project for his ongoing contribution in providing an updated list of apps in the Google Play Store.

\bibliographystyle{abbrv}
\bibliography{google-play} 

\end{document}